\documentclass[intlimits,twoside,a4paper]{article}

\usepackage[utf8]{inputenc}

\usepackage[eqsecnum]{cmpj3}

\articletype{A part of the Special collection to the 100th anniversary of the birth of Ihor Yukhnovskii}

\issue{2025}{28}{4}{43503}
\doinumber{10.5488/CMP.28.43503}
\title[3d random Ising model with LR]%
{Critical exponents of the Ising model with quenched structural disorder and long-range interactions at spatial dimension $d=3$}
\author[D. Shapoval, M. Dudka]{D. Shapoval\orcid{0000-0003-1648-2035}\refaddr{label1,label2}\thanks{Corresponding author: \email{shapoval@icmp.lviv.ua}.},
        M. Dudka\orcid{0000-0001-6971-8895}\refaddr{label1,label2,label3}}
\addresses{
\addr{label1} Yukhnovskii Institute for Condensed Matter Physics of the National Academy of Sciences of Ukraine, 79011 Lviv, Ukraine 
\addr{label2}  ${\mathbb L}^4$  Collaboration \& Doctoral College for the Statistical Physics of Complex Systems, Leipzig-Lorraine-Lviv-Coventry, Europe
\addr{label3} Lviv Polytechnic National University, 79013 Lviv, Ukraine
}

\date{Received November 19, 2025, in final form November 24, 2025}

\begin{document}

\maketitle

\begin{abstract}
We analyse the critical properties of a weakly diluted (random) Ising model with the long-range interaction decaying with distance $x$ as $\sim x^{-d-\sigma}$ in a $d$-dimensional space. It is known to belong to a new long-range random universality class for certain values of the decay parameter $\sigma$. Exploiting the field-theoretic renormalization group approach within the minimal subtraction scheme,  we compute the three-loop {renormalization} group functions. On their basis, with the help of asymptotic series resummation methods, we estimate the correlation length critical exponent $\nu$ characterising the new universality class for $d=3$ and for those values of $\sigma$ for which long-range interactions are relevant for the critical behaviour.

%
%
\printkeywords
%
\end{abstract}

\section{Introduction}


This year marks the centenary of Ernst Ising's paper \cite{Ising}, which has introduced the  nowadays famous Ising model. 
The model was suggested to Ernst~Ising by his adviser Wilhelm Lenz in order to  describe a ferromagnetic phase transition. Ernst~Ising  found no ferromagnetic ordering in one-dimensional chain of scalar magnetic moments (spins) with interaction between nearest neighbours for finite temperature~\cite{Ising}. However, since then, the  generalization of the Ising model for higher dimensions  and its modifications has been used  to explain a  collective behaviour in various systems from a  wide  range of scientific fields  both within and beyond the natural sciences \cite{Holovatch2026}. The critical properties of the Ising model were also in the focus of interests of the brilliant Ukrainian physicist Ihor Yukhnovskii, whose hundredth anniversary is also celebrated this year. Especially he was interested in the effects of the interaction range on the cooperative behaviour \cite{Yukhnovskii2001}. 
Authors of this paper studied a lot from his books \cite{Yukhnovskii2001,Yukhnovskii1995}.  Therefore, it is a great honor for us  to dedicate our paper on the Ising model to the memory of  Prof. Ihor~Yukhnovskii.

Our study concerns the analysis of the changes of critical behaviour of Ising model caused by mutual effects of two factors: long-range (LR) interactions and quenched disorder. The influence of a range of microscopic interactions between system components on critical behaviour is one of the most important problems in the theory of critical phenomena. Studies of LR interactions are highly significant due to their prevalence in physical, chemical, and biological systems \cite{review_a, review_b, lectures0, lectures}. Such interactions are usually modelled as algebraically  decayed with distance $x$ between interacting particles according to the power law
\begin{equation}\label{interaction}
J(x)=\frac{1}{x^{d+\sigma}},
\end{equation}
where $d$ stands for space dimension.
Models with interactions given by  (\ref{interaction}) have found wide applications in  different fields of physics \cite{review_a, review_b, lectures0, lectures}. 
Recent advances in technology for realizing controllable LR interactions in optical lattices, molecular and atomic systems  \cite{Monroe2021,Bernien2017}, particularly with the possibility of continuous tuning of the decay parameter $\sigma$  \cite{Yang2019}, have further increased interest in the theoretical study of systems with interactions of type (\ref{interaction}).

Models with LR  interactions possess distinctive features that affect their critical behaviour. Further on, we focus on the case of weak LR interactions, for which the decay parameter is positive, $\sigma> 0$. For this case, the one-dimensional Ising model ($d=1$) with $\sigma<1$ in contrast to its short-range (SR) counterpart,  was proven to exhibit a  phase transition to the long-range-ordered phase at non-zero temperature \cite{Dyson1969}. Field-theoretical renormalization group (RG) analysis of the LR
interacting $n$-vector model  (Ising system is its particular case  at $n=1$) reveals that there  are three universality classes depending on $\sigma$: (i) for $\sigma\leqslant d/2$ the classical mean-field critical behaviour is realised, (ii)  for $\sigma\geqslant 2$ critical exponents coincides with those of the model with SR interactions, (iii) for $d/2<\sigma<2 $ there is the LR universality class, for which critical exponents depend on $\sigma$ \cite{Fisher1972}. Later it was established that the boundary $\sigma^*$ between the SR and the LR universality classes is actually defined by a pair correlation function critical exponent of the SR model $\eta_{\rm SR}$, $\sigma^*=2-\eta_{\rm SR}$ \cite{Sak}. Such a picture was corroborated by other approaches including non-perturbative RG (NPRG) \cite{Defenu2015},
Monte Carlo simulations \cite{Luijten2002} and conformal bootstrap \cite{Behan2019} (for other references  see
\cite{Defenu2020}). The estimates characterizing the LR universality class have been obtained within Monte Carlo simulations for the Ising model mainly in spatial dimensions $d=1$ and $d=2$ (for collection of references see \cite{Benedetti2020}). Only a few Monte Carlo results are available for $d=3$ \cite{Belim2016,Belim2018}. RG results for these exponents  are available in low-loop approximations  \cite{Fisher1972,Suzuki1972,Belim2003}. The study within the highest accessible three-loop approximation was performed only recently \cite{Benedetti2020}  with a correction of loop-integral calculations this year \cite{BenedettiAddendum}.

Another factor discussed in this paper is the  impact of structural  disorder on the critical behaviour. Since some degree of  structural imperfection is unavoidably  present  in almost all materials, this issue still attracts interest of researchers.  It is known that the changes introduced by the disorder depend on its type: various realizations of quenched disorder have been studied for SR case (see, e.g., \cite{Holovatch02}).  However,  only two types of disorder have been considered for LR interacting systems in the form of dilution \cite{Yamazaki1978,Li1981,Theumann1981,Belim2003a,Chen2001,Chen2002,Shapoval2022,Shapoval2024} and in the form of random fields \cite{Affonso2023,Ding2024,Okuyama2025,Weir1987,Agrawal2023,Tarjus2020}. 

In its turn, LR interacting models with random field have been investigated by a bunch of approaches. 
In particular, different aspects of low-temperature long-range order have been explored by rigo\-rous~\cite{Affonso2023,Ding2024,Okuyama2025} as well as RG methods~\cite{Bray1986,Weir1987}. Dynamical ordering has been analysed numerically by Monte Carlo simulations \cite{Agrawal2023}.
The issue of possible dimensional reduction has been addressed within NPRG approach~\cite{Tarjus2020}. 

By contrast, the diluted case has been studied only by field-theoretical RG methods \cite{Yamazaki1978,Li1981,Theumann1981,Belim2003a,Chen2001,Chen2002,Shapoval2022,Shapoval2024}.  
These studies demonstrated relevance of  quenched dilution for critical behaviour and the  emergence of disorder-induced universality class.
Due to the degeneracy of RG functions for the Ising model with uncorrelated disorder at the one-loop level, the universal characteristics of this universality class have been obtained in the form of $\sqrt\epsilon$-expansions  \cite{Theumann1981,Chen2001,Chen2002,Shapoval2024}. These expansions are known to have bad convergence properties in the context of the study of SR models (see, e.g., \cite{Kompaniets2021}). There has been only one attempt to get reliable results for fixed space dimension in this case: resummation of the RG functions was performed within
a massive renormalization scheme \cite{Belim2003a}.
Almost all previous estimates have been obtained  within the two-loop approximation only,  while $\sqrt\epsilon$-expansion extracted from three-loop approximation \cite{Shapoval2024} was based on the RG  data with errors \cite{Benedetti2020}, corrected only recently \cite{BenedettiAddendum}.
Therefore, in our study we complete the  existing data  by our estimate of correlation length critical exponent based on resummation of three-loop RG functions.

The structure of the paper is as follows. In the next section~\ref{Sec2}, we introduce the model and its field-theoretical formulation. In section~\ref{Sec3}, we describe the renormalization procedure used and present RG functions. We treat these functions in  section~\ref{Sec4}  to extract values of the critical exponent $\nu$ at spatial  dimension $d=3$  with the help of resummation procedures. In section~\ref{Sec5} we summarize our findings with conclusions.

\section{Model}\label{Sec2}

\begin{figure}[t]
	\begin{center}
		\includegraphics[width=.2\paperwidth]{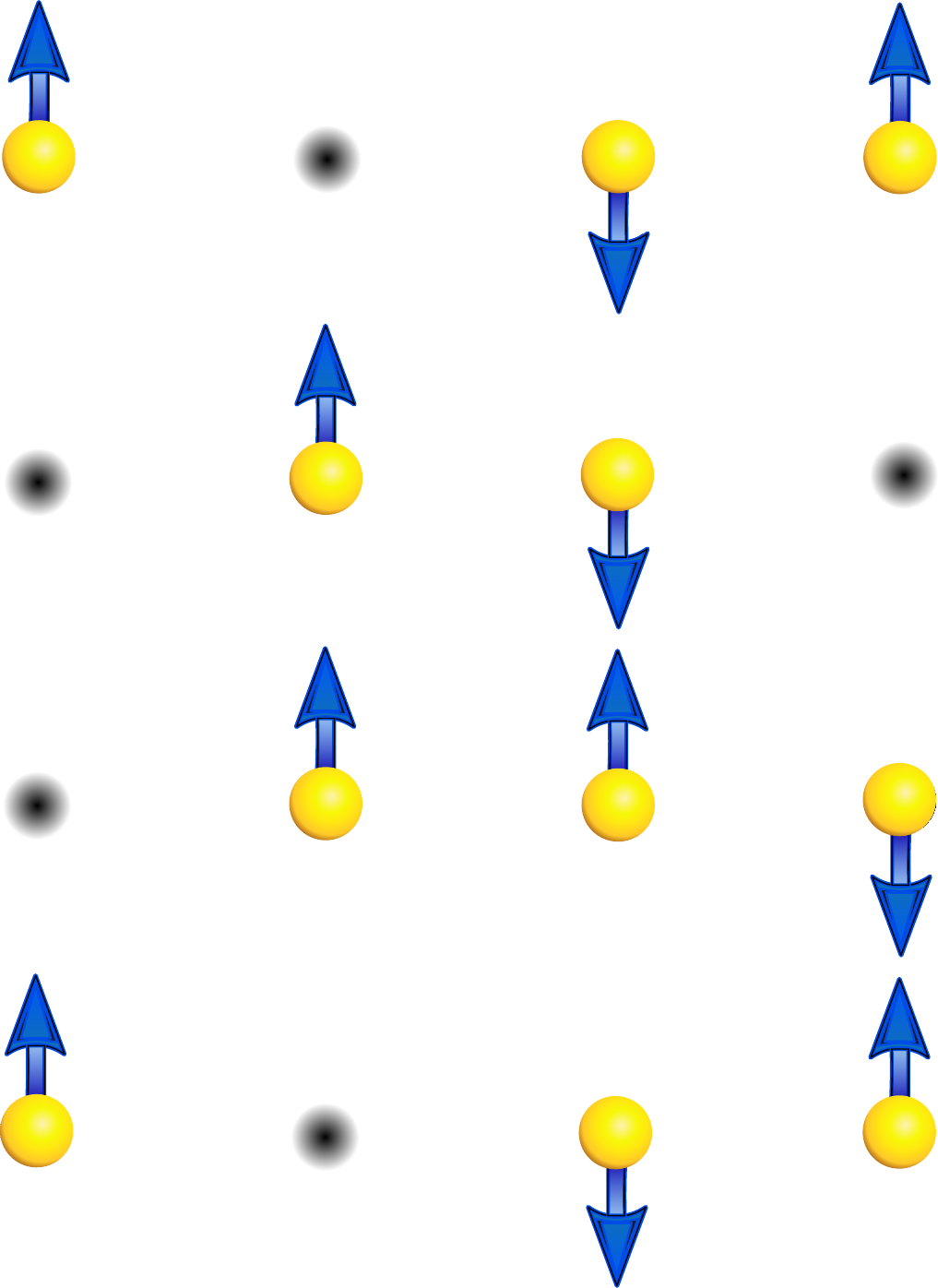}
	\end{center}
	\caption{(Colour online) Structurally disordered two-dimensional ($d = 2$) diluted Ising model: yellow disks represent lattice sites ${\bf x}$ occupied by Ising spins ${S}_{\bf x}$ (blue arrows), while dark gray sites correspond to non-magnetic impurities.}
        \label{dilution_fig}
\end{figure}

The macroscopic spin Hamiltonian 
of our model can be written in the following form:
\begin{equation}
  \label{ham_spin_dis}
      {\cal H} = - \sum_{{\bf x},{\bf x}'} J (|{\bf x}-{\bf x}'|) c_{{\bf x}} c_{{\bf x}'}  {S}_{\bf x} 
 {S}_{{\bf x}'},
\end{equation}
where ${\bf x}$ runs over site of $d$-dimensional (hyper)cubic lattice, $J (|{\bf x}-{\bf x}'|)>0$ is the ferromagnetic interaction between $S=\pm 1$ ``classical'' Ising spins on sites ${\bf x}$ and ${\bf x}'$, its form is defined by equation~(\ref{interaction}), and the occupation numbers $c_{{\bf x}}$ either equal to 1 if the lattice site ${\bf x}$ is occupied by a spin or equal to $ 0$ if site ${\bf x}$  is vacant. Here, we are interested in the case of the so-called weak dilution, when concentration of the occupied sites $c$  is much larger than percolation threshold. We consider that all $c_{\bf x}$ are independent and identically distributed (i.i.d.) random variables distributed according to
\begin{equation}\label{distr}
{\cal P} (\{c_{{\bf x}} \})=\prod_{\bf x} p(c_{{\bf x}}),
\qquad p(c_{{\bf x}})= c \delta (c_{{\bf x}}-1) + (1-c)
\delta (c_{{\bf x}}).
\end{equation}
We are interested in the case of quenched disorder, where the positions of vacancies/nonmagnetic impurities $c_{\bf x}$ are fixed (see figure~\ref{dilution_fig}). The {static and dynamic} critical behaviour of such random-site Ising model with  interactions between neighbour spins has been studied only by numerous Monte Carlo simulations (see e.g., reviews in \cite{Folk03,Kompaniets2021}). Unfortunately, we are unware of any studies of such a model when interactions decay with distance according to equation~(\ref{interaction}).

For systems with quenched disorder, the
 thermodynamics of physical observables can be analysed  from the
configurational average of the free energy \cite{Brout59}:
 \begin{equation}\label{av_free}
 F=-\beta^{-1} \overline{\ln Z (\{c\})}  
\, ,
\end{equation}
where $\beta^{-1} =k_\mathrm{B} T$, $k_\mathrm{B}$ is the Boltzmann constant and the configuration-dependent partition function reads
\begin{equation}
Z (\{c\})={\rm Sp} \, {\rm e}^{-\beta {\cal H}}, \hspace{2em}
{\rm Sp} \, (\dots) = \prod_{\bf{x}} \sum_{S_{\bf{x}}=\pm 1}\, (\dots) \, .
\label{2.3}
\end{equation}
Overline in (\ref{av_free}) means average over distribution (\ref{distr})
\begin{equation}
\overline {\left.\dots\right.}  = \prod_{\bf x} \sum_{c_{\bf x}}p(c_{\bf x})(\dots)\, .
\end{equation}
To avoid averaging of the logarithm in equation~(\ref{av_free}) one can utilize the replica trick \cite{Emery1975,Edwards1975,Dotsenko01}
\begin{equation} \label{2.14}
 \ln Z (\{c\}) = \lim_{m\to 0} \frac{\big[Z (\{c\})\big]^m -1}{m}\,.
\end{equation}
The $m$th power of the configuration-dependent partition function is represented in the form of a functional integral using the Stratonovich--Hubbard transformation. This transformation introduces field variables $\vec{\phi}^{\alpha}_{{\bf x}}$, ($\alpha=\{1,\dots, m\}$), conjugated to the spin variables. Subsequently, the trace over the spin subsystem is taken, and after configurational averaging, the free energy is obtained (see e.g.,~\cite{Holovatch_lecture}):
\begin{equation} \label{5.24}
{ F} \sim \int({\rm d}\vec{\phi}) \re^{-{\cal H}_{\rm eff}}, \qquad \int ({\rm d} {\phi}) = \prod_{\alpha=1}^n \prod_{\bf{x}} \int_{-\infty}^{\infty} {\rm d} {\phi}^\alpha_{\bf x}\, .
\end{equation}
The effective Hamiltonian in (\ref{5.24}) for the Ising  model with quenched weak dilution (\ref{ham_spin_dis})  and  with LR interactions  (\ref{interaction})   reads:
\begin{equation}
   \label{ham_eff_dis_lr}
        {\mathcal{H}}_{\rm eff}{=} \int \rd^d x \left\{ \frac{1}{2} \left[\tau |{\phi}^\alpha|^{2} {+} |\nabla^{\sigma/2} {\phi}^\alpha|^{2}\right] {+} \frac{u_0}{4!}\sum_{\alpha = 1}^{m} |{\phi}^{\alpha}|^{4} {+} \frac{v_0}{4!}\bigg[ \sum_{\alpha = 1}^{m}({\phi}^{\alpha})^{2}\bigg]^2\right\}\, .
\end{equation}
In the limit  $m\rightarrow 0$, the field theory (\ref{ham_eff_dis_lr}) describes the critical properties of a LR interacting random Ising model. The parameter $\tau$ is connected with the temperature distance to the critical point. The bare coupling $u_0$ is proportional to the concentration $ c$ and is positive whereas the bare coupling $v_0$ is proportional to $ c(c-1)$  and thus is negative. The last term in (\ref{ham_eff_dis_lr}) is present only for non-zero dilution: it is directly responsible for the fluctuations effective interaction due to the presence of impurities. { We remind that~(\ref{ham_eff_dis_lr}) was obtained with the assumption of a weak quenched dilution (far above the percolation threshold).} Note that theory (\ref{ham_eff_dis_lr}) is valid only for $\sigma < 2 - \eta_{\rm SR}$. Since we are interested only in the long-distance properties of the theory, in the limit of a small wave vector $k\to 0$, we kept only the first  non-trivial term of the expansion of the Fourier image of interactions $k^\sigma$.

\section{RG approach}\label{Sec3}

Within the field-theoretical RG approach \cite{Kleinert2001,Amit2005,ZinnJustin2021}, the analysis of the critical properties of $\phi^4$ models is based on the calculation of the  so-called one-particle irreducible vertex functions. These functions are perturbatively calculated around the Gaussian model [$u = v = 0$ in equation~(\ref{ham_eff_dis_lr})] in the coupling constants $u$ and $v$. However, these calculations result in integrals that are divergent at the upper critical dimension, which in our case is $d_u=2\sigma$. To remove these divergencies, a controlled rearrangement of the perturbative series for the vertex functions is performed by imposing certain normalizing conditions.  In what follows, we use the massless minimal subtraction scheme \cite{tHooft1972,tHooft1974}. 
We  express bare parameters via the renormalized ones, introducing the renormalizing factors of the field $\phi$, $Z_\phi$, the couplings, $Z_u$ and~$Z_v$ and for operator $\phi^2$, $Z_{\phi^2}$ ensuring finiteness of two-point vertex function with one $\phi^2$ insertion.

Then, the scaling bahaviour can be analysed from RG $\beta$-  and $\gamma$-functions  characterizing the change of the vertex functions under the RG transformation. They  are introduced  as follows: 
\begin{eqnarray}\label{RGf}
\beta_{u} = \frac{\partial u}{\partial \ln {\mu}}, \quad \beta_{v} = \frac{\partial v}{\partial \ln {\mu}},\quad \gamma_{\phi} = \frac{\partial Z_{\phi}}{\partial \ln {\mu}},  \quad \overline{\gamma}_{\phi^{2}} = - \frac{\partial \overline{Z}_{\phi^{2}}}{\partial \ln {\mu}}, 
\end{eqnarray}
where $\overline{Z}_{\phi^{2}} = Z_{\phi^{2}} Z_{\phi}$  and the derivatives are taken at fixed bare parameters. Here, ${\mu}$ is  the scale parameter introduced in the same way as in \cite{Benedetti2020}.

The critical region is controlled by a stable fixed point (FP) of the RG transformations, which must be also accessible from the initial conditions. The FPs are determined as solutions of the system:
\begin{eqnarray}
    \label{beta_dis}
        \beta_{u} (u^{*}, v^{*}) = 0,  \,\quad\,
        \beta_{v} (u^{*}, v^{*}) = 0.
\end{eqnarray}
The stability of the FPs (\ref{beta_dis}) is defined by the stability matrix 
$
    \left(  \begin{smallmatrix}
    {\partial \beta_{{u}}}/{\partial u}      &  {\partial \beta_{{u}}}/{\partial v} \\
    {\partial  \beta_{{v}}}/{\partial u} &  {\partial \beta_{{v}}}/{\partial v} 
\end{smallmatrix} \right)
$.
If this matrix, evaluated at a FP $(u^*,v^*)$, has both its eigenvalues $\lambda_{1}$ and $\lambda_{2}$ with positive real parts, then this FP is stable.

The critical exponents are calculated at the stable FP $(u^*,v^*)$ as follows:
\begin{equation}\label{cre}
\eta=2-\sigma+\gamma_{\phi}(u^*,v^*),\qquad
\nu^{-1}=2-\gamma_{\phi}(u^*,v^*) +\bar\gamma_{\phi^2}(u^*,v^*).
\end{equation}
All other critical exponents can be calculated from the scaling relations \cite{Kleinert2001,Amit2005,ZinnJustin2021}.

We work in the three-loop approximation, using the results of loop integral calculations \cite{Benedetti2020,BenedettiAddendum}. Note that our procedure differs. The authors of references~\cite{Benedetti2020,BenedettiAddendum} fixed spatial dimension $d$ and looked for poles in the deviation of the decay parameter from the real space dimension $\sigma=(d-\epsilon)/2$, while we fix $\sigma$ and look for poles in the deviations of spatial dimension from the upper critical dimension $d=2\sigma-\epsilon$.
The result of our calculations of the RG functions (\ref{RGf}) within the three-loop approximation is as follows:
\begin{eqnarray}\label{bu}
  \beta_u&=&u\Big[-\epsilon+\frac{3 u}{2}+2 v- \frac{1}{18} \left(27
   u^2+72 u v+42 v^2\right) D_\sigma\nonumber\\&&{}+\frac{1}{96}  \left(261 u^3+960 u^2 v+1154 u v^2+424 v^3\right) D_\sigma^2+\frac{1}{144} (81 u^3 + 324 u^2 v + 378 u v^2 + 136 v^3) E_\sigma\nonumber\\&&{}+\frac{1}{288}{ \left(81 u^3+288 u^2 v+378 u v^2+136 v^3\right)}F_\sigma+ \frac{1}{72}(3 u+4 v) \left(3 u^2+6 u v+2
   v^2\right)G_\sigma\nonumber\\&&{}+\frac{1}{216} (3 u+2 v) \left(9 u^2+30 u v+28 v^2\right)H_\sigma\Big],\\ \label{bv}
   \beta_v&=&v\Big[-\epsilon + 6 u+\frac{4}{3} v-\frac{1}{24}  \left(9 u^2+36 u v+22 v^2\right) D_\sigma\nonumber\\&&{}+\frac{1}{24} \left(21 u^3+94 u^2 v+129 u v^2+48 v^3\right)D_\sigma^2+\frac{1}{108} v  \left(63 u^2 + 111 u v + 44 v^2\right)E_\sigma\nonumber\\&&{}+\frac{1}{{72}} \left(9 u^3+30 u^2 v+45 u v^2+16 v^3\right)F_\sigma+ \frac{1}{108}(3 u+4 v) \left(3 u^2+6 u v+2
   v^2\right)G_\sigma\nonumber\\&&{}+\frac{1}{108} v  \left(9 u^2+24 u v+11 v^2\right)H_\sigma\Big],\\ \label{gf2}
   \bar\gamma_{\phi^2}&=&\frac{1}{2}  u+\frac{1}{3} v-\frac{1}{12}  \left(3 u^2+6 u v+2 v^2\right)D_\sigma\nonumber\\&&{}+ \frac{1}{864}\left(378 u^3+1125 u^2 v+990 u v^2+220 v^3\right)D_\sigma^2+\frac{1}{432} v \left(9 u^2+18 u v+4 v^2\right)E_\sigma \nonumber\\&&{}+\frac{1}{{\color{black}96}}  \left(6 u^3+19 u^2 v+18 u v^2+4
   v^3\right)F_\sigma+\frac{1}{216}(3 u+2 v) \left(3 u^2+6 u v+2 v^2\right) G_\sigma, \\
   \gamma_{\phi}&=&0. \label{gf}
\end{eqnarray}
{The expression (\ref{gf}) is exact since $Z_\phi=1$, because LR term ($k^\sigma$) does not renormalize
via its nonanalyticity in momentum spase \cite{Fisher1972,Sak}. Therefore,  pair correlation critical exponent is $\eta=2-\sigma$ for the cases where LR interactions are relevant for critical behaviour. Such a result for $\eta$ is supported by other approaches (see review in \cite{Defenu2020}).} Here,  $\epsilon=2\sigma-d$ and notations in (\ref{bu})--(\ref{gf2}) are as follows:
$D_\sigma=\psi(1)-2 \psi \left(\frac{\sigma }{2}\right)+\psi (\sigma )$, 
${E_\sigma=\psi ^{(1)}(\sigma )-\psi ^{(1)}\left(\frac{\sigma }{2}\right)}$, 
${F_\sigma=\psi^{(1)}(1)- \psi ^{(1)}(\sigma )}$, 
$G_\sigma=\frac{\Gamma
   \left(-{\sigma }/{2}\right) \Gamma (\sigma )^2 }{\Gamma \left({3 \sigma }/{2}\right)}$; where $\Gamma(\dots)$ is the gamma function, while $\psi^{(i)}(\dots)$ is the digamma function of $i$-order. Results for $H_\sigma$ were obtained only numerically {(for details in computation and numerical estimation for certain $\sigma$, see appendix~\ref{AppA})}. At the two-loop order (first lines of equations~(\ref{bu})--(\ref{gf2})) we reproduce the known RG results  up to normalization constants \cite{Chen2001}.

\begin{figure}[htbp]
	\begin{center}
		\includegraphics[width=.3\paperwidth]{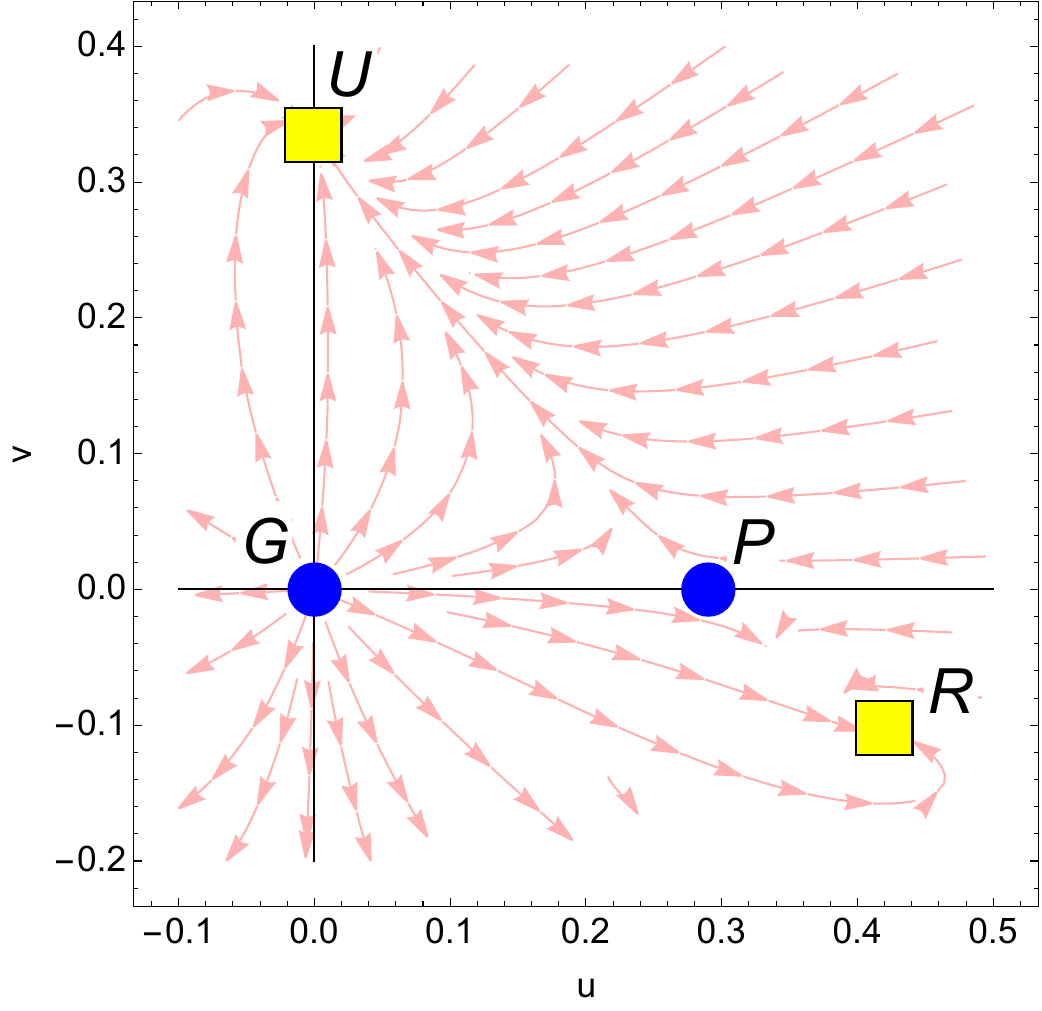}
	\end{center}
	\caption{{(Colour online) Schematic RG flows (denoted by red arrows) for the LR interacting random Ising model at $d<2\sigma$ in the parametric space of coupling constants $(u,v)$. Unstable FPs $G$ and $P$ are shown by blue discs, while stable FPs $U$and $R$ are depicted with yellow squares.}}
        \label{fixFP}
\end{figure}

Similarly to the SR case \cite{Folk03}, the $\beta$-functions (\ref{bu}), (\ref{bv}) for LR models with quenched dilution
are also known to have four FPs $(u^*,\,v^*)$ in the parameter space $(u, v)$ \cite{Theumann1981,Shapoval2022,Shapoval2024}. 
The physically relevant FPs are those that correspond to the initial conditions of the effective Hamiltonian (\ref{ham_eff_dis_lr}), namely $u > 0$ and $v < 0$. For our case in dimension $d<2\sigma$, we have the following picture (see figure~\ref{fixFP}):
(i) unstable Gaussian FP ${ G}$ ($u^{*} = 0$, $v^{*} = 0$); (ii) stable unphysical FP ${ U}$ ($u^{*} = 0$, $v^{*} \neq 0$), but  it is inaccessible from the physical initial conditions $u > 0$ and $v < 0$ set by (\ref{ham_eff_dis_lr});  (iii) unstable pure (system without impurities) FP ${ P}$ ($u^{*} \neq 0$, $v^{*} = 0$); (iv) stable random (disorder-induced) FP ${ R}$ ($u^{*} \neq 0$, $v^{*} \neq 0$). Therefore, only FP $R$ corresponds to the critical point of the LR interacting Ising model with quenched dilution. The calculation of the critical exponent $\nu$ for this fixed point is presented in the next section.

\section{Correlation length critical exponent $\nu$ }\label{Sec4}

\subsection{Resummation procedure}

Usually, there are two ways  to analyse the equations for the FP (\ref{beta_dis}) with $\beta$-functions of the form~(\ref{bu}),~(\ref{bv}): either developing the $\epsilon=2\sigma-d$ expansion, or directly by fixing spatial dimension and the decay parameter $\sigma$ and considering the renormalized couplings as the expansion parameters. 

We cannot exploit the usual $\epsilon$-expansion to obtain FP $R$, since the system of equations for FP is degenerated at the one-loop level. Therefore, we use here the second approach. Perturbative series in the coupling constants obtained within RG theory for universal quantities are known to be asymptotic at best rather than convergent, exhibiting factorially growing coefficients and thus a zero radius of convergence~\cite{Kleinert2001, Pelissetto2002}. Therefore, direct substitution of the couplings into the RG equations does not yield physically meaningful results, and an appropriate resummation procedure is required. The simplest approach is to apply Pad{\'e} approximants directly to the divergent series $f(z) = \sum_{n=0}^{L} f_{n} z^{n}$, representing them as ratios of polynomials $[M/N](z)$, whose coefficients  are fixed to reproduce the first $L = M+N$ known terms~\cite{BakerBook}. However, Pad{\'e} approximants often develop spurious poles on the positive real axis, preventing their use in physical regions. A more robust strategy is to employ the Borel--Leroy transformation \cite{Holovatch2001}:
\begin{eqnarray}\label{Bim}
B(z) = \sum_{n=0}^{L} \frac{f_n}{\Gamma(n+b+1)} z^n. 
\end{eqnarray}
Parameter $b$  is the Leroy parameter controlling the damping of large-order coefficients.
Then, the initial series is reproduced with
\begin{eqnarray}\label{invB}
f(z) = \int_0^{\infty} \re^{-t}\, t^{b}\,B(z t)\, \rd t.
\end{eqnarray}
In the Pad\'e--Borel--Leroy resummation procedure  Borel image (\ref{Bim}) in (\ref{invB}) is replaced by  its Pad\'e approximant. In our calculation we use  Pad\'e-approximant $[L-1/1](z t)$ with linear denominator:
\begin{eqnarray}\label{resB}
f^{\rm res}(z) = \int_0^{\infty} \re^{-t}\, t^{b}\,[L-1/1](z t) \, \rd t.
\end{eqnarray}
For $b = 0$, (\ref{resB}) reduces to the standard Pad{\'e}--Borel resummation \cite{Baker1976, Baker1978}. Let us note that any converged observable should be independent of artificially introduced parameters, such as the Leroy parameter. However, at any finite loop order $L$, the resummed series acquire an artificial dependence on such parameters. To extract a meaningful result, we assume that the optimal estimate of an observable at order~$L$ corresponds to the values of the artificial parameters for which the resummed quantity varies most weakly --- that is, where it becomes stationary with respect to the changes in these parameters. This criterion is known as the Principle of Minimal Sensitivity \cite{frustrated_b,frustrated_c}. Nevertheless, even the Pad{\'e}--Borel--Leroy method may fail for multivariable series due to the poles lying on the integration contour or due to poor convergence of Pad{\'e} approximants. 

Generalization of the Pad\'e--Borel--Leroy procedure on the case of several variables is obvious with the help of a resolvent series in an auxiliary variable. For instance, writing any function $f(u,v)$ as $f(u z, v z)$, one can perform procedure (\ref{Bim})--(\ref{resB}) for the series in $z$, eventually setting $z=1$ after all steps.

\subsection{Results at fixed dimension $d=3$}

 Let us start the discussion of the values of the correlation length critical exponent $\nu$ for the physically relevant case $d=3$, considering the two-loop RG functions (first lines of equations~(\ref{bu})--(\ref{gf2})). We provide the intermediate technical details in the appendix~\ref{AppC}.  Applying the Pad\'e--Borel--Leroy resummation to the two-loop $\beta_u$ and $\beta_v$ functions, we are able to found the FP $R$ by fixing $d=3$ for all values $\sigma$ between $3/2$ and $2-\eta_{\rm SR}$, where we use the result {$\eta_{\rm SR}=0.0362$} calculated for the SR Ising model in the highest accessible loop order \cite{Kompaniets17}. {At $\sigma=3/2$ (the case of upper critical dimension for three dimensional case, $\sigma=d/2$) the logarithmic corrections to the scaling are expected in the same way as in SR theory at $d=4$, although we do not touch this issue in our study.}

Substituting the obtained FP coordinates into the correspondingly resummed expression for the critical exponent $\nu$,  we obtain the data shown with red solid line  in figure~\ref{final}. The results are obtained for the optimal Leroy parameter, $b_{\mathrm{opt}} = 27$ (see appendix~\ref{AppC}). The dependence of $\nu$ on $\sigma$ is monotonic with a local maximum at $\sigma\approx 1.55$. With an increase of $\sigma$, the values of $\nu$ go towards the value of the correlation length critical exponent of the SR random Ising model, $\nu_{\rm RSR}$. We present its value $\nu_{\rm RSR}=0.665$, obtained applying the same Pad\'e--Borel--Leroy resummation to the corresponding SR RG functions, restricted at two-loop level, by  the dashed red line in figure~\ref{final}. The value  $\nu_{\rm RSR}=0.675$ calculated from RG functions accessible in the highest loop-approximation \cite{Kompaniets2021} is shown with the orange dot-dashed line. 

For comparison, we present in figure~\ref{final} the two-loop results for pure (undiluted) LR Ising model by dot-dashed line. They are obtained by extracting data for FP $P$ by solving the equation $\beta_u(u^*,0)=0$ with the correspondingly resummed  function $\beta_u$ and substituting into the expression for  $\nu$. The significant difference between the critical exponents of the random LR Ising model and those of its undiluted counterpart is observed. Therefore, it is interesting to verify this finding using another approach.

The previously used simple resummation techniques, which gave results for two-loop RG functions, turn out to be insufficiently reliable when applied to the fixed points derived from the three-loop RG functions (see discussion in the appendix~\ref{AppC}).   However, we cannot use more sophisticated methods based on conformal mapping \cite{frustrated_b,frustrated_c}, since the series for our RG functions obtained in low-loop approximations are two short. 
 
 \begin{figure}[htbp]
    \centering
    \includegraphics[width=0.45\textwidth]{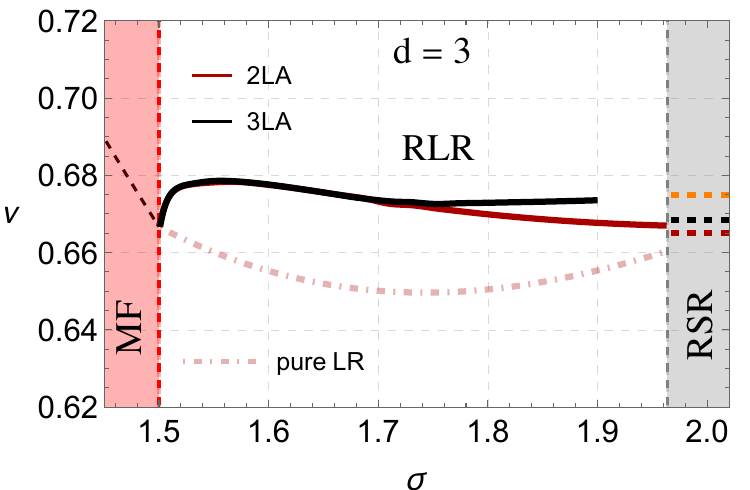}
    \caption{(Colour online) Correlation length exponent $\nu$ for the Ising model with LR interactions and quenched disorder at $d = 3$ as a function of $\sigma$. The data in the region $d/2<\sigma<2-\eta_{\rm SR}$ (denoted as~RLR) are obtained within the two-loop approximation with the help of the Pad{\'e}--Borel--Leroy procedure at $b_{\mathrm{opt}} = 27$  (red solid curve) and within the three-loop approximation with the help of the ``truncated'' version of the subsequent resummation procedure (solid black curve). The values of $\nu$ for the mean field~(MF) region ($\sigma{ \leqslant}  d/2$) are shown by thin dashed line. Estimates of $\nu$ for the random SR (RSR) model, valid for $\sigma<2-\eta_{\rm SR}$, obtained  within different approximations, are also shown: two-loop approximation value $\nu_{\rm RSR} = 0.665$  (red dashed line),  three-loop approximation value $\nu_{\rm RSR} = 0.668$  (black dashed line), six loop approximation value $\nu_{\rm RSR} = 0.675$ \cite{Kompaniets2021}. The light red dot-dashed curve presents two-loop data for the critical exponent $\nu$ of the undiluted LR Ising model.}
    \label{final}
\end{figure}
 
 Therefore, we employ a subsequent resummation procedure which has been shown to perform well in studies of the three-dimensional SR Ising model with weak quenched disorder. This method is designed for the resummation of a two-variable asymptotic series arising in the RG analysis of systems with quenched disorder \cite{Alvarez2000, Blavatska2001, Blavatska2003}. The technique involves the consecutive application of the Pad{\'e}--Borel--Leroy transformation: first with respect to the coupling associated with the pure system ($u$), and then with respect to the disorder-induced coupling ($v$), thereby implementing the principle of ``inequality of rights''~\cite{Pelissetto2000} between variables (for more details, see appendix~\ref{AppB}). This asymmetrical resummation procedure may help to avoid Pad{\'e} poles and provides stable estimates for the fixed points and their stability within high-order perturbative expansions \cite{Blavatska2001, Blavatska2003}. The most reliable fixed point estimates extracted from the three-loop RG functions are provided by the ``truncated'' version of the subsequent resummation procedure, which remains more stable under variations of the control parameter $\sigma$ (black solid line in figure~\ref{final}).

As one can see, for small values of $\sigma<1.75$, the three-loop results are very close to the two-loop values. With a further increase of $\sigma$, a larger deviation of the $\nu$ values from different approximations is observed. After $\sigma>1.9$ the three loop data for $\nu$ abruptly go down, which does not allow to obtain a reliable estaimate. 
Such behaviour raises the question of whether the two-loop approximation should be considered the optimal truncation for the random LR Ising model, or even of the validity of the perturbative RG approach in this case.

\section{Conclusion}\label{Sec5}

In this work, we investigated how the critical behaviour of an Ising system changes under the mutual influence of two factors: LR interactions and structural disorder. 
It is known that the random Ising model with LR belongs to a new ``random'' universality class, distinct from the pure (undiluted) LR universality class, as well as from the random SR universality class for the values of decay parameter $d/2<\sigma<2-\eta_{\rm SR}$. 

Aiming at providing a quantitative description of the critical behaviour in this new universality class, we considered a scalar  $\phi^4$ field theory with two coupling constants and employed the field-theoretic RG  approach.  We have derived the three-loop RG functions within massless minimal subtraction scheme. On the basis of these functions, we have calculated the correlation length critical exponent $\nu$ at a fixed spatial dimension $d=3$ and for the values of decay parameter $d/2<\sigma<2-\eta_{\rm SR}$ employing various techniques for analysing asymptotic series.

Our two-loop estimates obtained with help Pad{\'e}--Borel--Leroy resummation methods demonstrate a non-monotonous dependence of $\nu$ on $\sigma$ when changing from the mean-field  boundary $d/2$ to the SR boundary $2-\eta_{\rm SR}$. There is a significant difference with the corresponding values of $\nu$ for the undiluted LR Ising model. Therefore, we hope that our study will motivate verification of this finding within Monte Carlo simulations.

Herein, within the three-loop approximation we have obtained  reliable estimates for $\nu$ using the subsequent resummation approach only for $\sigma<1.9$. A possible explanation of this outcome is that the perturbative field-theoretical RG approach is not proper for the random Ising model with LR interactions, making the two-loop approximation to be optimal truncation in this case.


\section*{Acknowledgements}

We thank Prof. Yu. Holovatch for having introduced us to the fascinating world of critical phenomena and phase transitions, and we acknowledge numerous useful discussions with him during the work on this topic. The work was supported by the National
 Research Foundation of Ukraine Project 2023.03/0099
``Criticality of complex systems: fundamental aspects
 and applications''.
 

\appendix

\section{Computation $H_\sigma$}\label{AppA}
\renewcommand{\thefigure}{A.\arabic{figure}}
\setcounter{figure}{0}

\begin{figure}[h!]
	\centering
	\includegraphics[width=0.5\textwidth]{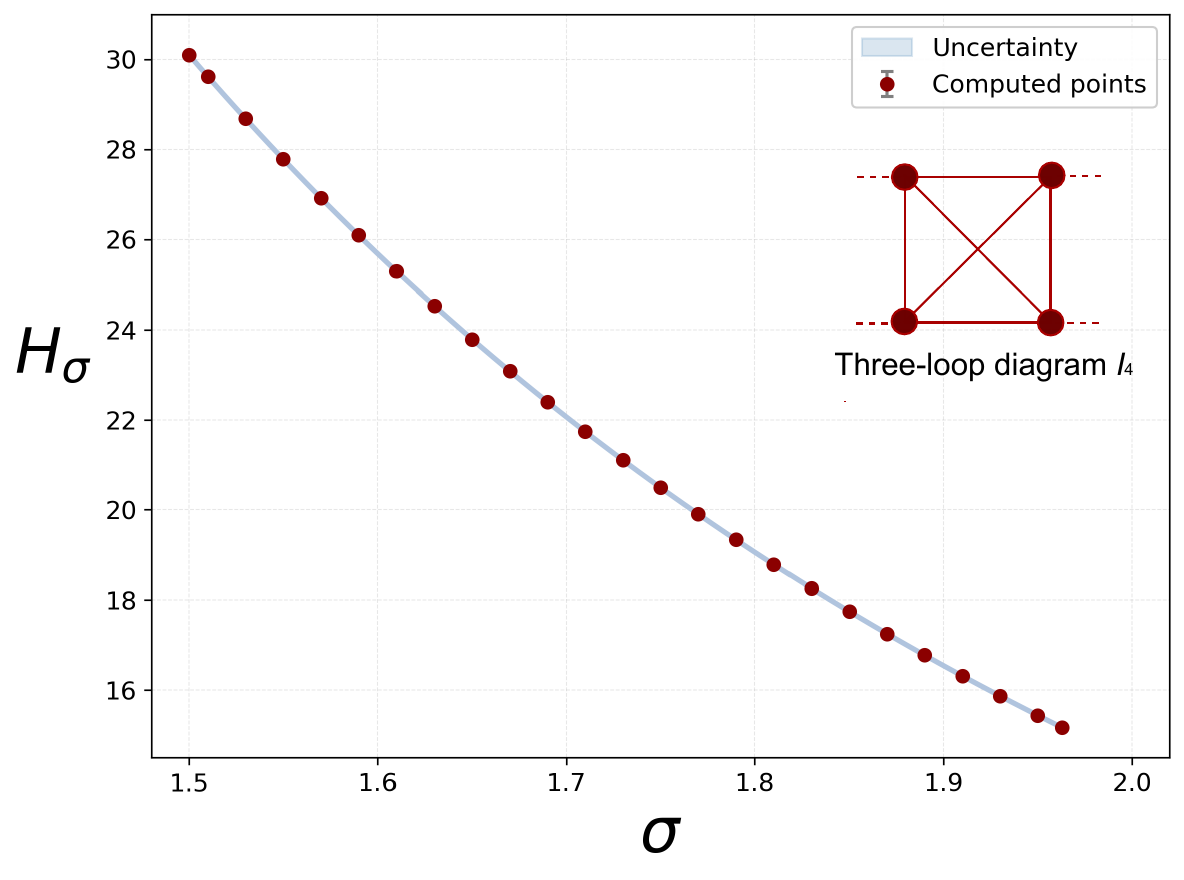}
	\caption{(Colour online) Numerical evaluation of $H_\sigma$ in equation~(\ref{f4sigma}) with respect to the control parameter $\sigma$. The numerical integration error is estimated as the difference between the results obtained at different levels of computational refinement and is determined by the extent to which the outcome stabilizes upon adding additional sampling points. The resulting uncertainty remains within the size of the plotted markers and does not affect the qualitative behaviour of the curves.}
	\label{num_f4}
\end{figure}

The contribution $H_\sigma$ in (\ref{bu}), (\ref{bv}) comes from the three-loop integral depicted by diagram $I_4$ in~\cite{Benedetti2020,BenedettiAddendum} (see inset of figure~\ref{num_f4}). As it was shown in \cite{BenedettiAddendum}  it reduces to calculation of  kite diagram.  Here, we employ the same method  \cite{BenedettiAddendum}, which is based on a decomposition into Hepp sectors, i.e., an ordering of Schwinger parameters that isolates $1/\epsilon$ divergences. Considering $I_4$ in the Schwinger-parameter representation, we obtain:
\begin{equation}
    \label{I40}
   I_4 \sim \int_{0}^{\infty} \rd \bar{a}_1 \ldots \rd \bar{c}_2 \frac{(\bar{a}_1 \bar{a}_2 \bar{b}_1 \bar{b}_2 \bar{c}_1 \bar{c}_2)^{\sigma/2 - 1} \re^{-(\bar{a}_1 + \bar{a}_2 + \bar{b}_1 + \bar{b}_2 + \bar{c}_1 + \bar{c}_2)}}{S^{d/2}}
\end{equation}
with
\begin{eqnarray}
    S &=& \bar{a}_1 \bar{a}_2 (\bar{b}_1 + \bar{b}_ 2 + \bar{c}_1 + \bar{c}_2) + \bar{b}_1 \bar{b}_2 (\bar{a}_1 + \bar{a}_2 + \bar{c}_1 + \bar{c}_2) + \bar{c}_1 \bar{c}_2 (\bar{a}_1 + \bar{a}_2 + \bar{b}_1 + \bar{b}_2) \nonumber \\
    &+& \bar{a}_1 (\bar{b}_2 \bar{c}_1 + \bar{b}_1 \bar{c}_2) + \bar{a}_2 (\bar{b}_1 \bar{c}_1 + \bar{b}_2 \bar{c}_2).
\end{eqnarray}
Here, as mentioned in \cite{BenedettiAddendum}, the denominator $S$ is symmetric by (i) exchange of the pairs $(\bar{a}_1, \bar{a}_2)$,$(\bar{b}_1, \bar{b}_2)$, $(\bar{c}_1, \bar{c}_2)$ and by (ii) simultaneous exchange of elements of two of the pairs. As a result, instead of the full $6! = 720$ sectors we have $(6!)/(3! 2^2) = 30$ distinct sectors need to be considered. 

Omitting the intermediate, rather cumbersome steps, we define $H_\sigma$, which in the case of interest $d \to 2 \sigma - \epsilon$, appears as the coefficient of the $1/\epsilon$-term in the following form:
\begin{eqnarray}
    \label{f4sigma}
   H_\sigma = 48 \frac{\Gamma{\left(\sigma\right)^{3}}}{\Gamma{\left(\frac{\sigma}{2}\right)^{6}}} \sum_{j = 1}^{30} \int_{0}^{1} \prod_{i = 1}^{5} \rd t_{i} \frac{(t_{1} t_{5})^{\sigma/2 - 1} (t_{2} t_{4})^{\sigma - 1} (t_{3})^{3\sigma/2 - 1}}{\tilde{S}_{j}^{\sigma}} ,
\end{eqnarray}
which can be evaluated numerically for specific values of $\sigma$ (see figure~\ref{num_f4}). Each $\tilde{S}_{j}$ in (\ref{f4sigma}) is the sector-specific polynomial that arises after (i) ordering the six Schwinger parameters into one of the 30 inequivalent sectors, (ii) performing the change of variables $\{\bar{a}_1, \bar{a}_2, \bar{b}_1, \bar{b}_2, \bar{c}_1, \bar{c}_2\} \to \{\bar{a}_1, t_1, t_2, t_3, t_4, t_5\}$ and integrating out $\bar{a}_1$.

\section{Two-loop resummation stability and three-loop resummation limitations}\label{AppC}
\renewcommand{\thefigure}{B.\arabic{figure}}
\setcounter{figure}{0}

\begin{figure}[h!]
    \centering
    \includegraphics[width=0.45\textwidth]{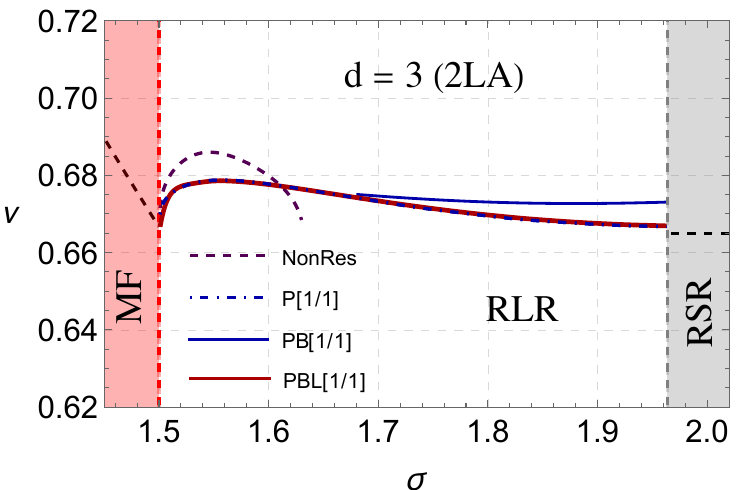}
    \caption{(Colour online) Correlation length exponent $\nu$ for the Ising model at $d = 3$ as a function of $\sigma$ in two-loop approximation. Dark purple dashed line: critical exponent before resummation; blue dot-dashed line: resummation using the Pad{\'e} approximant; blue solid line: the Pad{\'e}--Borel resummation; while red solid line: the Pad{\'e}--Borel--Leroy resummation.}
    \label{2LA}
\end{figure}

Let us restrict the system of equations (\ref{bu})--(\ref{gf2}) to the two-loop perturbative order and for numerical evaluation of the two-loop correlation length critical exponent $\nu$ we use the resummation techniques described in section~\ref{Sec4}. As can be seen in figure~\ref{2LA}, reliable results can be obtained even with the simplest Pad{\'e} approach (blue dot-dashed line). However, our analysis shows that the Pad{\'e}--Borel resummation breaks down in a certain range of the control parameter $\sigma$ (blue solid line). By contrast, the Pad{\'e}--Borel--Leroy procedure remains stable (red solid line). Using the Principle of Minimal Sensitivity, we determine the optimal value $b_{\rm opt} = 27$, at which the fixed point values become stationary with respect to variations of $\sigma$. 

\begin{figure}[h!]
    \centering
    \includegraphics[width=0.45\textwidth]{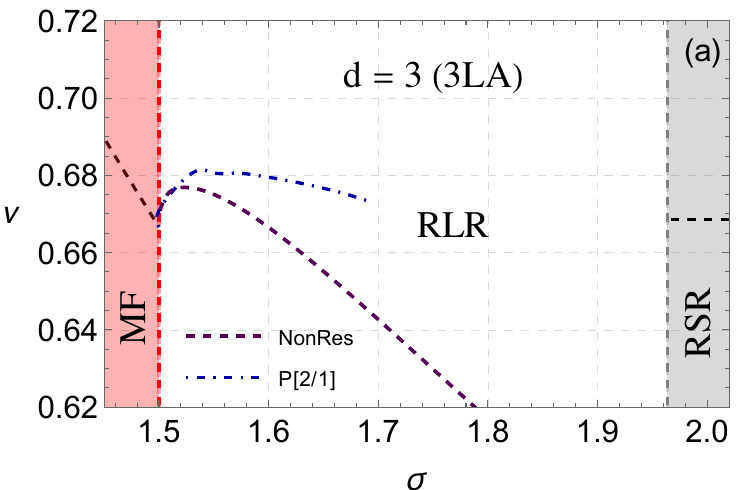}
    \includegraphics[width=0.45\textwidth]{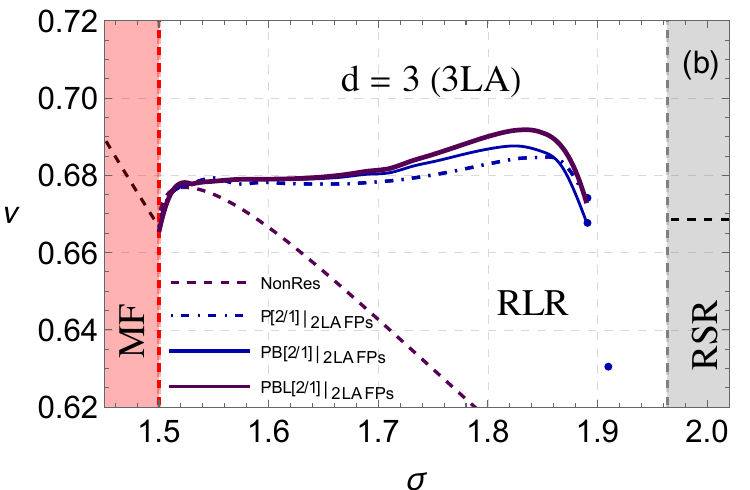}
    \caption{(Colour online) Correlation length exponent $\nu$ for the Ising model at $d = 3$ as a function of $\sigma$ in three-loop approximation calculated in FPs from: (a) three-loop RG $\beta$-functions and (b) two-loop RG $\beta$-functions, correspondingly. Dark purple dashed line: critical exponent before resummation; blue dot-dashed line: resummation using the Pad{\'e} approximant; blue solid line: the Pad{\'e}--Borel resummation; while dark purple solid line: the Pad{\'e}--Borel--Leroy resummation.}
    \label{3LA}
\end{figure}

In comparison with the two-loop results obtained using the same resummation techniques, we are unable to effectively  extract reliable estimates of the three-loop critical exponents [see figure~\ref{3LA} (a)]. While Pad{\'e} approximation still shows intervals of regular behaviour separated by the region of controlled parameter where the resummed expression diverges, Pad{\'e}--Borel and Pad{\'e}--Borel--Leroy fail more severely, producing manifestly unphysical results. This indicates that, at the three-loop level, more sophisticated resummation strategies are required to obtain meaningful critical exponents.

A common strategy for handling asymptotically divergent series is to determine the fixed point from a resummed RG $\beta$-function at a lower loop order, and then substitute this value into the expressions for the critical exponents calculated at a higher order \cite{LeGuillou1980, Kleinert2001, Pelissetto2002}. This approach typically yields more stable estimates, since $\beta$-function series are usually less well-behaved, whereas the accuracy of the critical exponents tends to improve when higher-order expansions are used \cite{LeGuillou1980, Kleinert2001, Pelissetto2002}. The estimate of the correlation-length critical exponent $\nu$ obtained within this approach is shown in figure~\ref{3LA} (b).

\section{Subsequent resummation technique}\label{AppB}
\renewcommand{\thefigure}{C.\arabic{figure}}
\setcounter{figure}{0}


In our case, the RG functions contain two coupling constants: the ``pure'' coupling $u$ and the ``disorder-induced'' coupling $v$. Consequently, the RG flow is governed by two functions, $\beta_u(u,v)$ and $\beta_v(u,v)$, each of which can be expanded as a double series in powers of $u$ and $v$ [see equations~(\ref{bu})--(\ref{gf2})]. To improve the resummation of multivariable series, a two-step (respectively, \emph{subsequent}) resummation    technique based on the  ``inequality of rights'' between the pure and random couplings  was proposed in the context of the zero-dimensional ($d = 0$) diluted SR Ising model \cite{Alvarez2000}.  This technique was then developed in the context of the three-dimensional counterpart \cite{Blavatska2001, Blavatska2003}.
 In this approach, the series are treated asymmetrically with respect to $u$ and $v$. 

Let us illustrate this procedure for a generic function $f(u,v)$:
\begin{eqnarray}
    f(u,v) = \sum_{k=0}^{p} A_k(u)\, v^k,
\qquad
    A_k(u) = \sum_{n=0}^{p-k} f_{n, k}\, u^n.
\end{eqnarray}
Each coefficient $A_k(u)$ is itself a divergent series in $u$. For each of them, we construct the Pad\'e--Borel--Leroy image and extrapolate it using a Pad\'e-approximant  $\big[\frac{p-k-r_k}{r_k}\big](u t)$ in the inverse transform:
\begin{eqnarray}
   A_k^{(\mathrm{res})}(u) =
    \int_0^{\infty} \re^{-t}\, t^{b_1}\, \left[\frac{p-k-r_k}{r_k}\right](u t)\, \rd t, 
\end{eqnarray}
where $b_1$ is the Leroy parameter for the $u$-series. 
At the calculations, we use $r_{k}=1$ for all $0\leqslant k\leqslant p $.
After this stage, the function $f(u,v)$ is rewritten as
\begin{eqnarray}\label{f1}
    f^{(1)}(u,v) = \sum_{k=0}^{p} A_k^{(\mathrm{res})}(u)\, v^k,
\end{eqnarray}
which is now a function of $v$ with resummed $u$-dependent coefficients.
The second resummation step is then performed with respect to $v$:
\begin{eqnarray}
    f^{(\mathrm{res})}(u,v) = 
        \int_0^{\infty} \re^{-s}\, s^{b_2}\, \left[\frac{p-1}{1}\right]_u(v s)\, \rd s ,
\end{eqnarray}
where $\big[\frac{p-1}{1}\big]_u(v s)$ is Pad\'e-approximant for (\ref{f1}) with denominator linear in $v$, and $b_2$ is the corresponding Leroy parameter for the $v$-series.

In practical calculations, the new parameter $q$ can be introduced to represent the number of terms in~(\ref{f1}) which are to be resummed. Two main variants of the subsequent resummation have been tested:

\begin{enumerate}
    \item[(i)] {Full resummation ($q=p$)}:
    the second-stage Borel--Pad{\'e}--Leroy transform is applied to the
    entire resummed series $\sum_{k=0}^{p} A_k^{(\mathrm{res})}(u)\,v^k$.
    This approach utilizes all available perturbative information but may be sensitive to Pad{\'e} poles on the positive real axis.

    \item[(ii)] {Truncated resummation ($q=p-1$)}:
    only the first $p-1$ coefficients are resummed with respect to $u$,
    while the last coefficient $A_p(u)$ remains unresummed.
    Thus, before the second-stage resummation one has
    \begin{eqnarray}
    f^{(1)}(u,v) =
    \sum_{k=0}^{p-1} A_k^{(\mathrm{res})}(u)\, v^k.
    \end{eqnarray}
    The final resummed function is then obtained as
    \begin{eqnarray}
    f^{(\mathrm{res})}(u,v) =
    \int_0^{\infty} \re^{-s}\, s^{b_2}\,
    \left[\frac{p-2}{1}\right]_u(v s)\, \rd s + A_p(u)\, v^p .
    \end{eqnarray}
    This ``truncated'' version usually avoids Pad{\'e} poles
    and provides more stable numerical re\-sults~\cite{Blavatska2001, Blavatska2003}.
\end{enumerate}

\bibliographystyle{cmpj}
\bibliography{references}

\ukrainianpart

\title{Критичні показники моделі Ізінга із замороженим структурним безладом та далекосяжними взаємодіями при вимірності простору $d = 3$}
\author{Д. Шаповал\refaddr{label1,label2}, М. Дудка\refaddr{label1,label2,label3}}
\addresses{
\addr{label1} Інститут фізики конденсованих систем імені І.Р. Юхновського Національної академії наук України, 79011 Львів, Україна 
\addr{label2} Спiвпраця ${\mathbb L}^4$ \& Докторський коледж iз статистичної фiзики складних систем, Ляйпцiґ-Лотарингiя-Львiв-Ковентрi, Європа
\addr{label3} Нацiональний унiверситет ``Львiвська полiтехнiка'', 79013 Львів, Україна
}

\makeukrtitle

\begin{abstract}
\tolerance=3000%
Ми аналізуємо критичні властивості слабко розведеної (випадкової) моделі Ізінга з далекосяжною взаємодією, що згасає з відстанню $x$ як $\sim x^{-d-\sigma}$ у $d$-вимірному просторі. Відомо, що за певних значень параметра згасання взаємодії $\sigma$ така система належить до нового далекосяжного випадкового класу універсальності. Використовуючи підхід теоретико-польової ренормалізаційної групи у схемі мінімального віднімання, ми обчислюємо трипетлеві реноргрупові функції. На їхній основі, із застосуванням методів пересумовування асимптотичних рядів, ми оцінюємо критичний показник кореляційної довжини $\nu$, що характеризує новий клас універсальності для $d=3$ та для тих значень $\sigma$, за яких далекосяжні взаємодії є релевантними для критичної поведінки.
\keywords{критичні явища, модель Ізінга, системи з безладом, далекосяжні взаємодії, ренормалізаційна група}

\end{abstract}

\end{document}